\documentclass[11pt]{article}
\usepackage{hyperref}
\usepackage{amsthm}
\usepackage{amsmath}
\usepackage{amsfonts}
\usepackage{amssymb}
\usepackage{graphicx}
\usepackage{natbib}
\usepackage{subfigure}
\usepackage{algorithm}
\usepackage{algpseudocode}
\usepackage[defaultcolor=black]{changes}

\hypersetup{colorlinks=true,
            linkcolor=blue,
            urlcolor=black, 
            citecolor=blue}

\usepackage{setspace}
\title{\vspace{-1cm}Actively Learning Joint Contours of Multiple Computer
 Experiments}
\author{Shih-Ni Prim\thanks{Department of Statistics, NC State University} 
  \and Kevin R. Quinlan\thanks{Lawrence Livermore National Laboratory}
  \and Paul Hawkins\footnotemark[2]
  \and Jagadeesh Movva\footnotemark[2]
  \and Annie S. Booth\thanks{Corresponding author: Department of Statistics,
    Virginia Tech, {\tt annie\_booth@vt.edu}}}
\date{\today}

\begin{document}

\maketitle

\begin{abstract} Contour location---the process of sequentially training a 
surrogate model to identify the design inputs that result in a pre-specified
response value from a single computer experiment---is a well-studied active
learning problem.  Here, we tackle a related but distinct problem: identifying
the input configuration that returns pre-specified values of multiple 
computer experiments simultaneously.  
Motivated by computer experiments of the rotational torques acting upon a 
vehicle in flight, we aim to identify stable flight conditions that 
result in zero torque forces.  We propose a ``joint contour
location'' (jCL) scheme that strikes a strategic balance between exploring the
multiple response surfaces while exploiting learning of the intersecting
contours.  \added{Rather than working exploration and exploitation into a single
acquisition function, we devise two distinct acquisition schemes with a decision 
rule to choose between the two, which also 
provides a natural stopping criterion if no solution is present.}
We employ \added{traditional Gaussian processes (GPs), multitask GPs, and deep
GPs,} but our jCL procedure is applicable to any surrogate that can provide posterior
predictive distributions.  Our jCL designs significantly outperform existing
(single response) CL strategies, \added{optimization-based alternatives, and 
previous strategies for targeting joint contours,} enabling us to efficiently 
locate the \added{optimal configurations} for our motivating computer experiments.
\end{abstract}


\section{Introduction}
\label{s:intro}

Computer simulation experiments are invaluable tools in many scientific
fields, but particularly in the study of flight dynamics where physical 
experimentation is impractical across the large design space. 
\citep[e.g.,][]{pamadi2004aerodynamic,vassberg2008development,economon2016su2,quinlan2024}.  
Here, our focus
is on identifying a joint contour of multiple computer experiments.
Specifically, let $f^{(r)}:\mathcal{X}\rightarrow\mathbb{R}$ denote the $r^\textrm
{th}$ computer experiment for $r=1,\dots,R$.  Each $f^{(r)}$ is deterministic,
\added{continuous}, and acts on the same domain 
$\mathcal{X}\subset\mathbb{R}^d$.  Our goal is to
identify the ``optimal design point'':
\begin{equation} \label{e:objective}
\left\{\tilde{\mathbf{x}}\in\mathcal{X} \;\big\vert\; f^{(r)}\left(
    \tilde{\mathbf{x}}\right) = \tau_r \;\,\forall\;\, r = 1, \cdots, R\right\},
\end{equation} where $\tau_r$ are pre-specified target response values.
\added{Throughout, we will focus on identifying a single $\tilde{\mathbf{x}}\in\mathcal{X}$,
although we will discuss cases where zero or multiple solutions exist.}

For example, we are
motivated by high-fidelity computational fluid dynamics (CFD)
simulations of a ``High-Speed Army Reference Vehicle'' (HARV) 
in flight \citep{vasile2022high}.
\added{Multiple CFD simulations return the rotational torques 
acting on the vehicle (i.e., roll, pitch, and yaw moments)
as functions of the orientation of the vehicle's fins (visualized later in Figure \ref{f:data})
and several flight condition parameters (Mach speed, angle of attack, angle of 
sideslip, and deflections of control surfaces).  Stable flight occurs
when all rotational torques acting on the vehicle are zero,
also known as the ``trim condition.''  Ultimately, we aim to compile an accurate trimmed
aerodynamic database containing fin orientations that result in stable flight
for various flight conditions.  In this work we consider fixed
positions of the top and bottom fins, and focus
on identifying the optimal configuration of the side fins that results in zero pitch 
and zero roll.  For a given sideslip angle, there will be only one such 
possible configuration.  Efficient identification of the optimal fin configuration 
will facilitate application across a wide variety of settings.  Given the dimensionality
of this motivating application with relatively few movable fins, we restrict our 
exercises to low dimensional settings and discuss extensions to larger dimensions
in Section \ref{s:discuss}.}

Identifying $\tilde{\mathbf{x}}$ is particularly challenging when simulations 
are computationally expensive and evaluation budgets are limited.  We seek a 
sample efficient method for identifying $\tilde{\mathbf{x}}$ from as few 
evaluations of the expensive $f^{(r)}$ as possible.  This task requires a 
surrogate: a statistical model trained to emulate a computer experiment from
limited training data.  Effective surrogates provide accurate predictions
at unobserved inputs with appropriate uncertainty quantification (UQ).  Gaussian
process (GP) surrogates are favored as they offer nonlinear regression 
with closed-form posterior predictive distributions in a Bayesian framework
\citep{santner2003design,rasmussen2006gaussian,gramacy2020surrogates}.
\added{Multitask GPs can incorporate correlations among multiple responses
\citep{bonilla2007multi}, which can improve performance if the $f^{(r)}$ 
are not independent.}
Recent advances have aimed to retain the desirable properties of GPs while
additionally incorporating nonstationary flexibility 
\citep{booth2024nonstationary}, with deep Gaussian processes 
\citep[DGPs;][]{damianou2013deep} jumping to the forefront 
\citep[e.g.,][]{rajaram2020deep,marmin2022deep,sauer2023deep,ming2023deep,yazdi2024deep},
\added{but multitask DGP surrogates have yet to be explored.  Crucially, our
active learning procedure is agnostic to the choice of surrogate (as long as 
posterior predictive distributions are available) and can leverage any of these GP
variants.}

When evaluations are limited, training data may be strategically collected through 
an iterative model-informed process called active learning (AL).  In contrast
to static one-shot experimental designs like Latin hypercube samples
\citep[LHS;][]{mckay2000comparison}, active learning alternates between selecting
new design points based on an existing surrogate and retraining the surrogate
with the newly collected data.  New design points are chosen by optimizing an 
acquisition function which objectively quantifies the utility of potential design 
points based on existing surrogate knowledge.  Acquisition criteria may be
tailored to particular objectives such as variance reduction
\citep[e.g.,][]{cohn1996active,binois2019replication,song2026efficient},
optimization \citep[e.g.,][]{jones1998efficient,pourmohamad2021bayesian},
calibration \citep[e.g.,][]{koermer2024augmenting,surer2025simulation},
or sensitivity analysis \citep[e.g.,][]{wycoff2021sequential,belakaria2024active}.

Our motivating problem is most akin to that of contour
location (CL).  Contour location is a type of active learning where the
acquisition function specifically targets a level set,
$\{\mathbf{x}\in\mathcal{X} \mid f(\mathbf{x}) = \tau\}$, \added{and is designed to
balance exploitation of the contour with exploration of the design space.  Previous
CL works have adapted common Bayesian optimization procedures 
\citep[e.g.,][]{ranjan2008sequential,picheny2010adaptive} or stepwise uncertainty
reduction strategies \citep[e.g.][]{bect2012sequential,chevalier2014fast,marques2018contour}
to target level sets instead of optima.  More recent works have employed variations
of classification entropy as acquisition criteria \citep[e.g.,][]{cole2023entropy,booth2025contour}.
While there is a vast body of literature on CL for single computer experiments,
methods targeting joint contours of multiple experiments are scarce.
\citet{graziani2024targeted} propose an adaptive design strategy for identifying
inputs of multiple stochastic computer simulations whose response values fall within
specified tolerances with sufficient probability.  Their acquisition function is
built upon Gaussian probability density functions (PDFs) centered at the target 
value for each response, but much of their machinery is inspired by the
high noise level of each simulation they consider.  They target ranges 
$[\tau_r - \epsilon_r, \tau_r + \epsilon_r]$ instead of precise $\tau_r$, and their
surrogates can benefit from replicate observations to inform noise estimation.
In our deterministic setting, we benefit from more aggressive targeting of $\tau_r$ 
and must intentionally discourage replicates.  We find the PDF-based acquisition
scheme to be too conservative for our settings (more on this in Section \ref{s:sims}).}

\added{In order to identify $\tilde{\mathbf{x}}$ in Eq.~(\ref{e:objective}) from as few
evaluations of $f^{(r)}(\mathbf{x})$ as possible,
we propose a novel acquisition scheme which we term ``joint contour 
location'' (jCL).  Our acquisition procedure involves two mechanisms: one that
aggressively exploits identification of the optimal design point and one that
promotes exploration of all response surfaces, with a decision rule to decide
which mechanism to trigger at each acquisition.  This contrasts previous works that
aim to merge exploration and exploitation into a single acquisition function.
When surrogate certainty in the localization
of $\tilde{\mathbf{x}}$ is low (which is common at the start of the design phase), 
we promote exploration of all response surfaces
by targeting the Pareto front of posterior predictive uncertainty across all 
$r=1,\dots,R$.  When surrogate certainty in the
localization of $\tilde{\mathbf{x}}$ is high,
we prioritize exploitation using the joint posterior probability of each 
$f^{(r)}(\mathbf{x})$ being within a yet unattained distance of $\tau_r$.  Our
exploitation strategy is designed to quickly find a single $\tilde{\mathbf{x}}$
even if multiple exist; we defer extensions to ensure exploration of multiple
equivalent solutions for future work.
Our ``decision tree'' framework provides a natural stopping
criterion if no solution exists: simply halt the design if too many exploration
steps have been triggered in a row without any progress towards exploitation.
Crucially, our procedure is agnostic to the choice of surrogate as long as
posterior predictive probabilities are available.  We will benchmark our methodology
against a variety of competing strategies using several GP-based surrogates to demonstrate
its applicability.}

The remainder of this manuscript is organized as follows.  Section 
\ref{s:review} reviews surrogate modeling essentials while setting up an
illustrative example.  Section \ref{s:method} details our jCL procedure.
We validate our methodology on several synthetic
exercises in Section \ref{s:sims} before deploying it on our motivating
problem in Section \ref{s:harv}.  Section \ref{s:discuss} concludes with 
discussion of relevant extensions of this work.

\section{Gaussian Process Fundamentals}\label{s:review}

Let $\mathbf{x}_i\in\mathcal{X}\subset\mathbb{R}^d$ denote a $d$-dimensional
design point.  For a single black-box function, denote scalar
output as $y_i = f(\mathbf{x}_i)$.  Let $\mathbf{X}_n$ denote the $n\times d$ matrix
of $n$ row-combined design points, and let $\mathbf{y}_n$ denote the corresponding
response vector.

A Gaussian process is a potentially infinite collection of random variables, any
finite subset of which is distributed as a multivariate Gaussian
distribution \citep{rasmussen2006gaussian}.  A standard Gaussian 
process prior on $f(\mathbf{x})$ assumes 
$\mathbf{y}_n \sim \mathcal{N}_n(\boldsymbol\mu_n, \Sigma(\mathbf{X}_n))$ for any
design points $\mathbf{X}_n$.  Without loss of generality, we assume $\boldsymbol\mu_n=0$ 
after centering responses.  The covariance matrix $\Sigma(\mathbf{X}_n)$ contains
elements $\Sigma(\mathbf{X}_n)^{(ij)} = k(\mathbf{x}_i, \mathbf{x}_j)$ for $i\in\{1,\dots,n\}$,
$j\in\{1,\dots,n\}$ where kernel $k$ determines the covariance between $y_i$ and $y_j$
based on $\mathbf{x}_i$ and $\mathbf{x}_j$.
Standard kernels, like the squared exponential and Mat\`{e}rn 
\citep{stein1999interpolation}, are inverse functions of Euclidean distance, 
encoding the assumption that inputs closer to each other are more likely to have 
similar outputs.  Throughout, we use $\Sigma(\mathbf{A}, \mathbf{B})$ to
represent the matrix with $ij^\textrm{th}$ element containing the covariance 
between the $i^\textrm{th}$ row of $\mathbf{A}$ and the $j^\textrm{th}$ 
row of $\mathbf{B}$.  We also use $\Sigma(\mathbf{A})$ as shorthand
for $\Sigma(\mathbf{A}, \mathbf{A})$.

Conditioned on observed $\{\mathbf{X}_n, \mathbf{y}_n\}$, the posterior predictive 
distribution at a new location $\mathbf{x}^\star$ follows
\begin{equation} \label{e:pred}
f_n(\mathbf{x}^\star) \mid \mathbf{X}_n, \mathbf{y}_n \sim 
    \mathcal{N}_1(\mu, \sigma^2) \;\;\text{where}\;\; 
    \begin{cases}
    \mu = \Sigma(\mathbf{x}^\star, \mathbf{X}_n)
        \Sigma(\mathbf{X}_n)^{-1}\mathbf{y}_n\\ 
    \sigma^2 = \Sigma(\mathbf{x}^*) - \Sigma(\mathbf{x}^*, \mathbf{X}_n) 
        \Sigma(\mathbf{X}_n)^{-1} \Sigma(\mathbf{X}_n, \mathbf{x}^*).
    \end{cases}
\end{equation}
Posterior probabilities over intervals may be obtained through the 
application of the standard Gaussian cumulative distribution function (CDF).

\begin{figure}[h!]
\centering
\includegraphics[width=0.7\textwidth, trim=20 20 20 20]{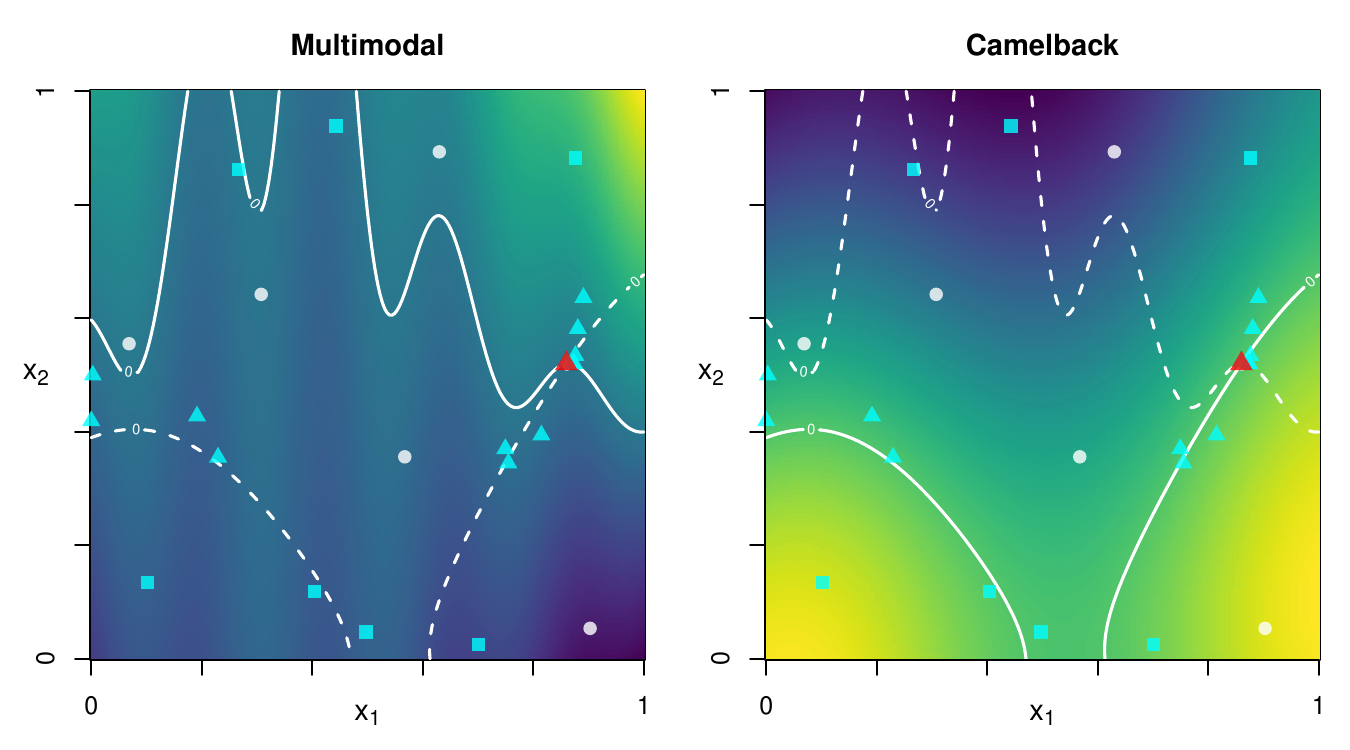}
\caption{
Heatmaps (yellow/high, purple/low) of the 2d multimodal function 
\citep[left;][]{bichon2008efficient} and the 2d camelback function 
\citep[right; derived from][]{molga2005test}. Solid white lines show the respective
contours $f^{(r)}(\mathbf{x})=0$.  Red triangle marks the optimal design
point where both contours are zero.  White circles indicate an initial LHS,
and cyan squares (exploration) and triangles (exploitation) indicate 
acquisitions made by our jCL scheme.}
\label{f:mm_cb}
\end{figure}

\begin{figure}[h!]
\centering
\subfigure[5 LHS points + 0 acquired points from jCL]{
  \includegraphics[width=.4\linewidth, trim=0 0 0 38, clip=True]{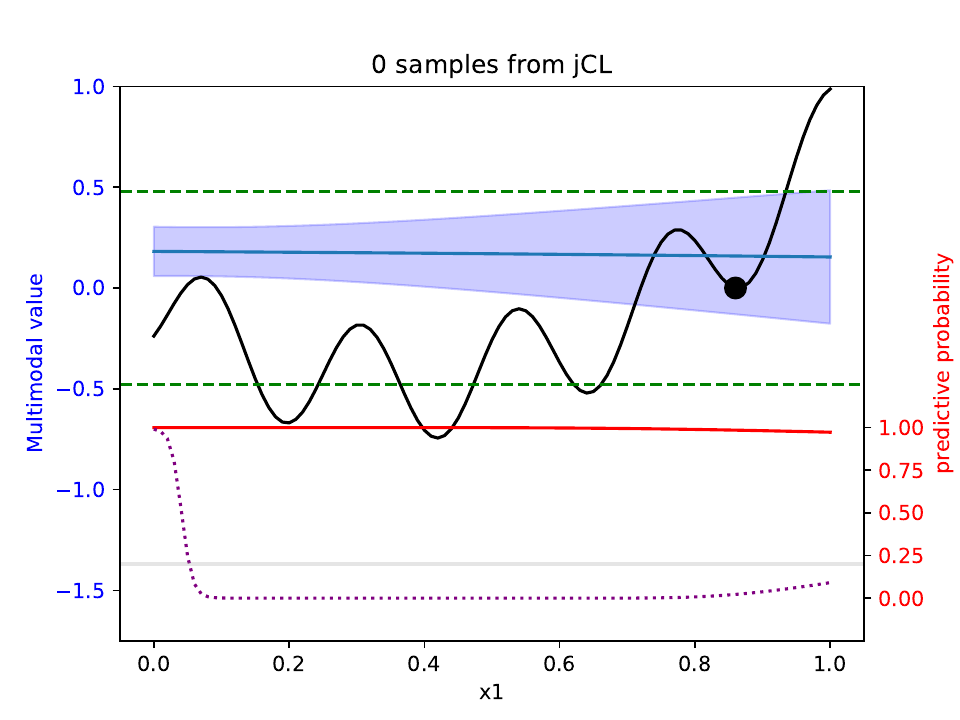}
  \includegraphics[width=.4\linewidth, trim=0 0 0 38, clip=True]{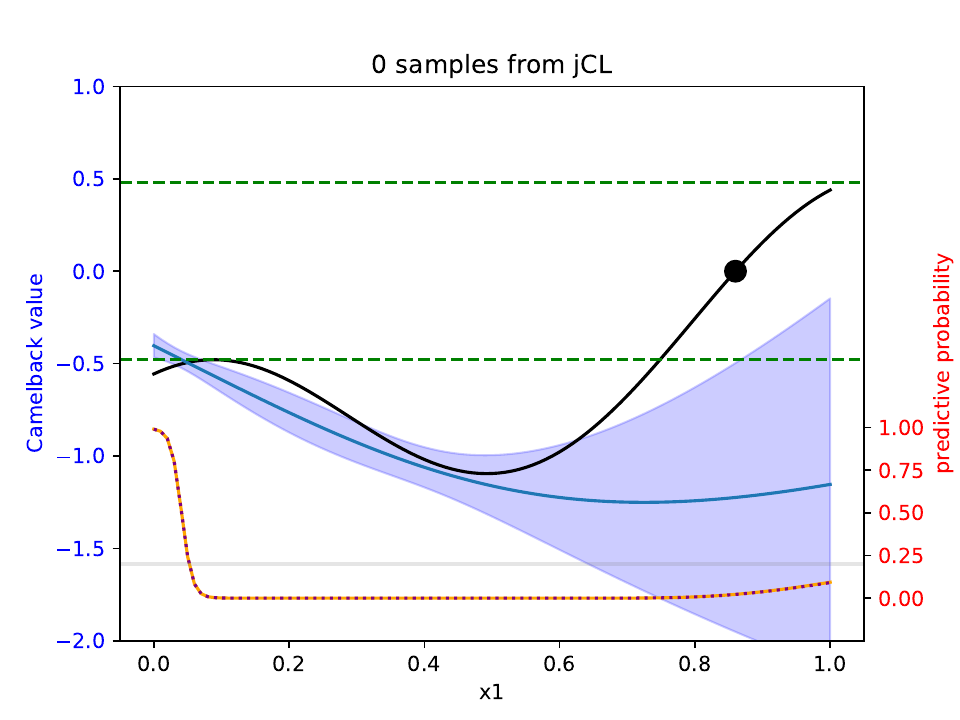}}
\subfigure[5 LHS points + 7 acquired points from jCL]{  \label{f:mm_cb_slice_b}
  \includegraphics[width=.4\linewidth, trim=0 0 0 39, clip=True]{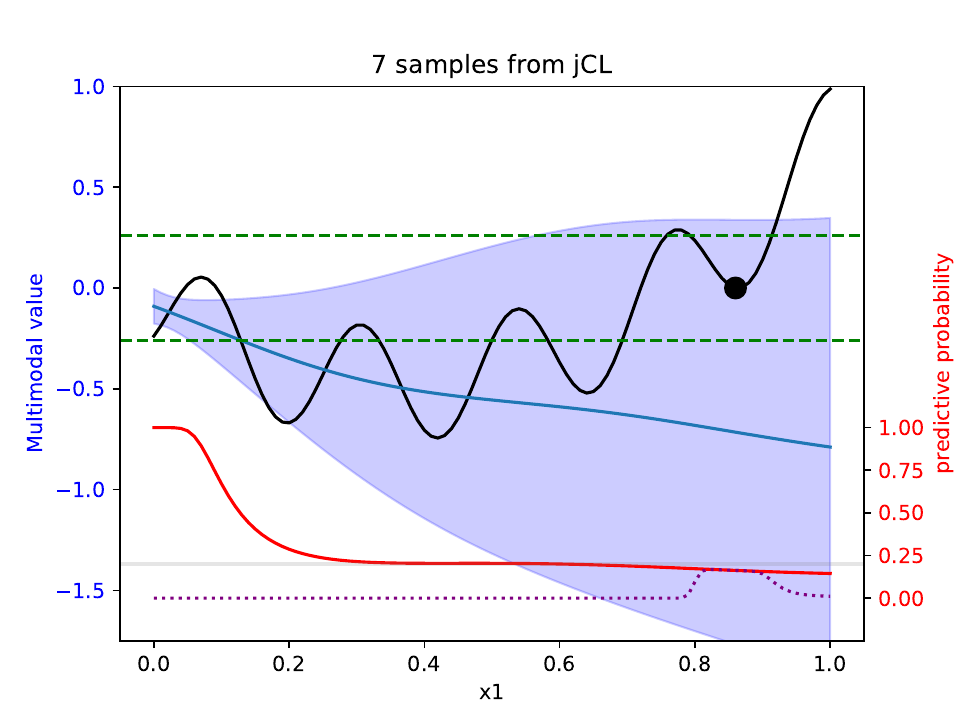}
  \includegraphics[width=.4\linewidth, trim=0 0 0 39, clip=True]{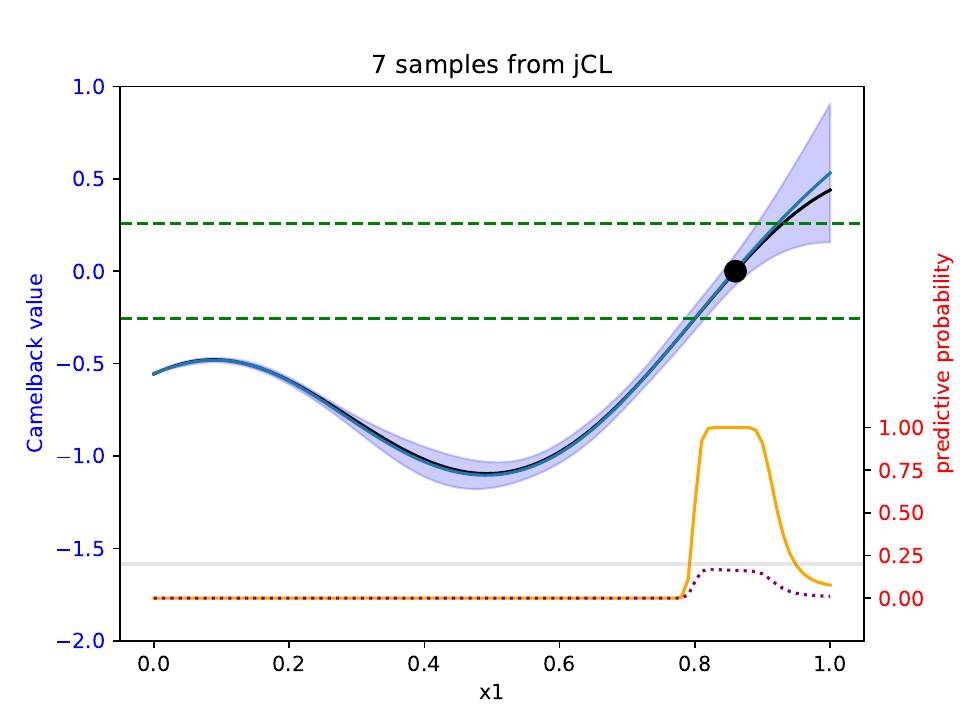}}
\subfigure[5 LHS points + 20 acquired points from jCL]{
  \includegraphics[width=.4\linewidth, trim=0 0 0 38, clip=True]{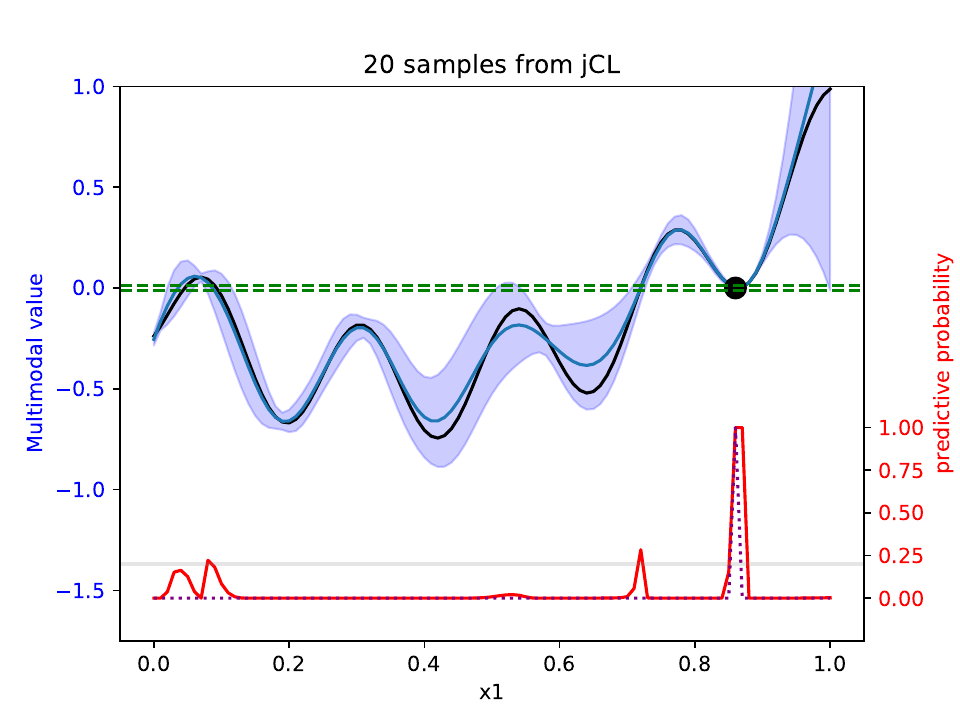}
  \includegraphics[width=.4\linewidth, trim=0 0 0 38, clip=True]{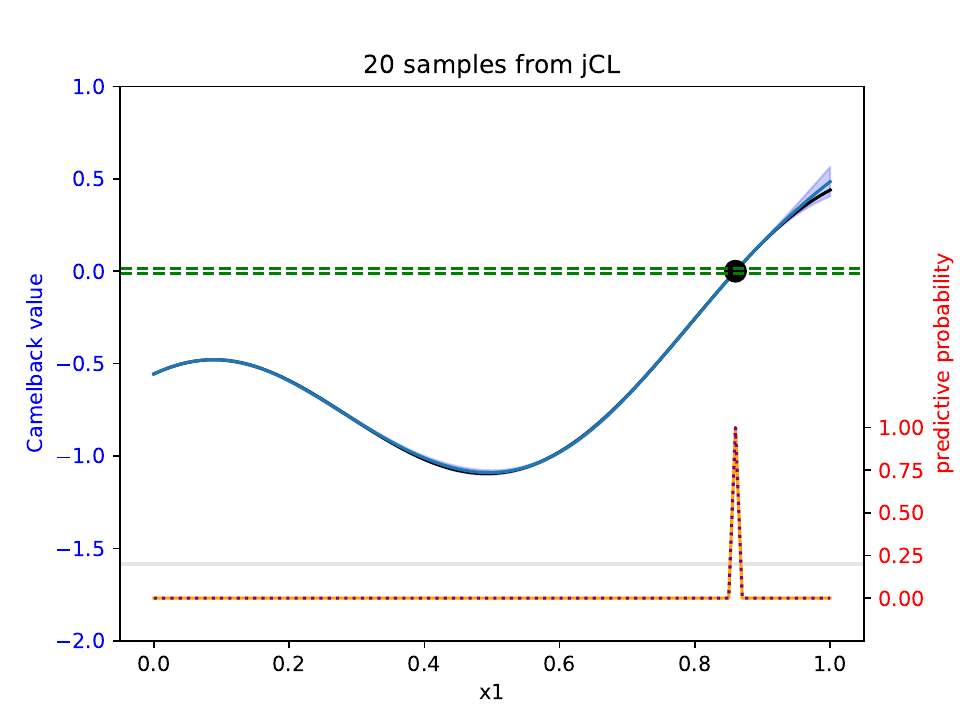}}
\caption{Slices at $x_2 = 0.52$ for the multimodal (left) and camelback (right) 
functions.  Black line shows true surface; black circle highlights $\tilde{\mathbf{x}}$
where both functions return zero.  Independent GP surrogates shown in blue
starting with a 5-point LHS and proceeding
with 0/7/20 jCL acquisitions across the two dimensions.
Horizontal green dashed lines mark the target ($\tau_1=\tau_2=0$) plus/minus the 
tolerance ($t_n$, Eq.~\ref{e:tol}).  Red (multimodal) and orange 
(camelback) lines along the $x$-axis show the separate probabilities 
$\mathbb{P}(-t_n < f^{(r)}(\mathbf{x}) < t_n)$, on a relative scale. Purple dotted lines show the 
resulting joint probabilities.  Joint probabilities that do not surpass
$p^\star=0.2$ (horizontal gray line) will trigger an exploration step.}
\label{f:mm_cb_slice}
\end{figure}

To demonstrate, consider the two-dimensional multimodal and camelback 
functions shown in Figure \ref{f:mm_cb} with $\tau_1 = \tau_2 = 0$
(after scaling).\footnote{Formulaic details of all
synthetic functions and domains are provided in Supplementary Material.}
We observed both functions at the same 5 input locations (the random
LHS indicated by the white circles), then trained independent GPs for 
each surface.  The upper panels of Figure \ref{f:mm_cb_slice} show the GP posterior means (solid blue)
and 95\% credible intervals (shaded blue) for each function 
along the slice $x_2=0.52$.  While we would not expect the surrogates
to perform well with only 5 observations, these figures still 
offer a useful visual of the GP's nonlinear regression and UQ capabilities.
We will revisit the other features of Figures \ref{f:mm_cb}--\ref{f:mm_cb_slice} 
in later sections.

When response surfaces are nonstationary---meaning the covariance between two 
observations depends on more than just the Euclidean distance between 
locations---deep Gaussian processes (DGPs) offer superior performance.
DGPs feature a ``hierarchical cascading of Gaussians'' \citep{dunlop2017howdeep},
in which outputs of one GP feed as inputs to another.  Inspired by spatial
warpings \citep{sampson1992nonparametric}, and akin to neural networks,
the latent layers of a DGP warp the input
space into a plausibly stationary regime, but their multidimensional functional
form poses a challenge to posterior inference.
To circumvent the difficult task of posterior integration, many have embraced
approximate variational inference 
\citep[e.g.,][]{damianou2013deep,salimbeni2017doubly,marmin2022deep}.
Yet when data sizes are limited and UQ is essential,
Bayesian approaches that use elliptical slice sampling \citep[ESS;][]{murray2010ESS}
to infer the latent warpings are preferable
\citep{sauer2023active,ming2023deep,sauer2023vecchia}.  The DGP's posterior 
distribution is conditionally Gaussian given
ESS samples of latent layers, which enables the calculation of 
posterior probabilities through careful application of the standard Gaussian
CDF \citep{booth2025contour}.

\added{When modeling multiple responses, it may be beneficial to incorporate 
correlations among $f^{(r)}(\mathbf{x})$ for $r=1,\dots,R$.  
Multitask GPs \citep{bonilla2007multi}, also 
called ``multi output GPs'' or ``vector valued GPs,'' expand the GP's 
prior covariance to include all pairings across locations ($i=1,\dots,n$)
and across functions ($r=1,\dots,R)$ \citep{alvarez2012kernels}.  The
posterior distribution is still Gaussian, and posterior
probabilities can again be calculated with careful application of Gaussian
CDFs.  Extensions
of multitask methods to deep Gaussian processes are still on the 
frontier; we are not aware of any publicly available software implementations.
Notably, our jCL active learning procedure
is plug-and-play with any surrogate that can provide posterior predictive
distributions.  We will demonstrate applicability to GPs, DGPs, and 
multitask GPs (which we will broadly refer to as ``GP-based surrogates'')
in Sections \ref{s:sims} and \ref{s:harv}.}

\section{Joint Contour Location} \label{s:method}

We seek an objective acquisition \added{procedure} that targets the optimal
design point $\tilde{\mathbf{x}}$, such as the red triangle indicated in
Figure \ref{f:mm_cb}, with as few evaluations of the expensive functions
as possible.  \added{Throughout, we consider evaluation of 
$f^{(r)}(\mathbf{x})$ for all $r=1,\dots,R$
as a single observation.  Any time we evaluate
a particular $\mathbf{x}$, we observe all responses at that same location.}  
Our objective contrasts existing CL methodologies that target 
an entire contour.  Figure \ref{f:mm_cb} visualizes this disconnect; 
spending expensive acquisitions to learn the entirety of each contour 
(solid/dashed white) is not the most efficient or effective way to pin 
down the optimal design point (red triangle).  

\added{We propose two 
distinct acquisition strategies---one targeting exploitation of the optimal
design point and one providing exploration of the response surfaces---with 
a decision rule to determine which to use for each acquisition.  In 
this section, we will introduce these
strategies separately before integrating them into one 
seamless AL procedure.  At each acquisition step, we presume to
have observed $n$ data points and have trained GP-based surrogates, 
$f_n^{(r)}$ for $r=1,\dots,R$.  Our objective is to determine which
input, $\mathbf{x}_{n+1}$, to observe next.}

\subsection{Exploitation} \label{s:exploitation}

A reasonable \added{exploitive} acquisition is any input which, with 
high probability \added{given existing surrogate knowledge},
could return $\tau_r$ for all $r=1,\dots,R$.  [The ideal 
acquisition would of course be the exact $\tilde{\mathbf{x}}$.]  Since
we are operating with a continuous response variable, we must consider
probabilities over intervals rather than probabilities of exact outcomes.
Let $t$ represent \added{some} tolerance value.  For
independent surrogates trained on $n$ observations, 
the joint posterior probability of a particular 
input residing within tolerance for all functions is
\begin{equation} \label{e:jointprob} 
J_n(\mathbf{x}, t) = \prod_{r=1}^R \mathbb{P}\left( 
    \tau_r - t \le f^{(r)}_n(\mathbf{x}) \le \tau_r + t\right).
\end{equation}
With \added{GP-based} surrogates, these probabilities involve Gaussian
CDF computations.  \added{This form may be easily modified for multitask
surrogates with correlations among responses.}  For simplicity, 
we use a constant tolerance across 
all functions after scaling each function to equivalent ranges, but
our methodology could easily accommodate unique $t_r$.

This ``joint probability'' (for short) is a natural 
acquisition function for a sequential design targeting the intersection 
of multiple contours.  Simply acquire
\[
    \mathbf{x}_{n+1} = \underset{\mathbf{x} \in \mathcal{X}}{\arg\max} \;\;
    J_n(\mathbf{x}, t).
\]
Yet the choice of tolerance $t$ can make-or-break this criterion.
If $t$ is too small, the $J_n(\mathbf{x}, t)$ criterion may be entirely flat
(with the surrogates not anticipating any inputs will meet this stringent
tolerance) or may be too peaky (with only a tiny portion of the input space
offering a nonzero probability), which could thwart standard optimizations.
If $t$ is too large, we run the risk of one of our observed data points
actually falling within tolerance across the board.
If an observed $\mathbf{x}_i$ for $i\in\{1,\dots,n\}$ satisfies 
$\tau_r - t \leq f^{(r)}(\mathbf{x}_i) \leq \tau_r + t$ for all $r=1,\dots,R$,
then, given the deterministic nature of our computer experiments, we will have
$J_n(\mathbf{x_i}, t) = 1$, which is the highest possible joint probability.
In that case, the maximum joint probability is guaranteed to reside at $\mathbf{x}_i$,
but $\mathbf{x}_i$ is already in our training data and would be a waste of an acquisition.

We propose an adaptive refinement of the tolerance which: \added{(1) gradually
converges to zero to guide acquisitions closer and closer to the joint contour
and (2) guarantees $J_n(\mathbf{x}_i, t)$ will be \added{approximately zero} for any 
observed $\mathbf{x}_i\in \mathbf{X}_n$ in order to 
discourage duplicate observations, which are wasteful in our deterministic setting.}  
Specifically, we set
\begin{equation} \label{e:tol} 
t_n = w\cdot\left[\min_{i=1,\dots,n} \left( \max_{1 \le r \le R} \left| y_{ir} - \tau_r \right| \right)\right] 
\quad\textrm{where}\quad y_{ir} = \added{f^{(r)}(\mathbf{x}_i)}
\quad\textrm{and}\quad 0<w<1.
\end{equation}
Acquisitions use $J_n(\mathbf{x}, t_n)$, which is adjusted as acquisitions are made
and $n$ is incremented.
Let's break this down from the inside out.  First, we take each observed response value
and find its distance from the contour.  For a particular observation, we pull the largest
of these distances across all functions.  This is a measure of the ``worst-case'' result for
a particular input.  For example, even if $\added{f^{(1)}(\mathbf{x})} = \tau_r$ exactly, if 
$\added{f^{(2)}(\mathbf{x})} = \tau_r + 100$, then the quantity 
$\max_{1 \le r \le 2} \left| y_{ir} - \tau_r \right|$ will equal 100.  Then, we pull
the minimum of all these ``worst-case'' distances across all the observed data points.
Think of this as the ``best of the worst.''  For our ``best'' observed data point 
(the one that got the closest to having all response values equal $\tau_r$), we 
grab the distance that was furthest off from all the functions within that data point.
Then, finally, we shrink that distance by a factor $w$ which must be strictly less than 1.
\added{This tolerance specification ensures any observed $\mathbf{x}_i\in\mathbf{X}_n$ will 
not be targeted since it will have near zero $J_n(\mathbf{x}_i, t_n)$ because 
$f_n^{(r)}(\mathbf{x})$ will equal $f^{(r)}(\mathbf{x})$
in our deterministic setting (as long as the surrogate properly preserves interpolation).  
Crucially, $t_n$ will only shrink as $n$ is incremented if a ``better'' point is 
acquired (meaning its $\max_{1 \le r \le R} \left| y_{r} - \tau_r \right|$ is lower than
the previous low).  While the joint probability surface is driven by surrogate predictions, 
the setting of $t_n$ is independent of the surrogate.  For continuous response surfaces,
this iterative shrinking of the tolerance will greedily guide acquisitions 
toward the optimal $\tilde{\mathbf{x}}$.}

\added{The scaling parameter $w$ is a user-adjustable tuning parameter that controls how
aggressively the tolerance is tightened after each ``better'' observation is found.
Values that are too large (i.e., $w\approx 1$) can slow progress.  Values that are too
small (i.e., $w < 0.5$) can cause the regions of non-zero joint probability to shrink 
too quickly, making the surface more difficult to optimize.  We find $w=0.9$ 
works well.  We use this default in all of our exercises, with further comparisons
reserved for Supplement \ref{a:tuning}.}  We also acknowledge
that variations of Eq.~(\ref{e:tol}) which leverage different distance metrics might
work equally well, although we favor the interpretability of the absolute value.

Returning to the illustrative example of Figure \ref{f:mm_cb_slice}, the dashed 
green lines show $\pm t_5$, $\pm t_{12}$, and $\pm t_{25}$ chosen according to 
Eq.~(\ref{e:tol}).  As the design progresses, the tolerance shrinks, honing in on the
contour.  The independent probabilities, 
$\mathbb{P}\left(\tau_r-t_n \le f^{(r)}_n(\mathbf{x}) \le \tau_r + t_n\right)$ for $r\in\{1,2\}$, 
are shown by the solid red/orange lines along the $x$-axis.  They capture the 
surrogate's belief that the response will fall within the tolerance bounds for a given
$\mathbf{x}$.  The joint probability formed by the product of these is displayed 
in dotted purple.  The next acquisition
would be at the peak of this joint probability (albeit in two dimensions -- the slice
is just for illustration).  Notice the joint probability focuses on regions where
both probabilities are high; it is not led astray by inputs that have high
probabilities for the multimodal function but not the camelback function, or vice versa.  
\added{Larger tolerances are necessary early on when the surrogates do not have
a lot of information regarding the location of $\tilde{\mathbf{x}}$.  As closer
observations are acquired, shrinking the tolerance facilitates effective exploitation
as the joint probability peak converges around the true
$\tilde{\mathbf{x}}$ (purple circle).}

\begin{figure}[h!]
\centering
\includegraphics[width=0.8\linewidth, trim=0 0 0 30, clip]{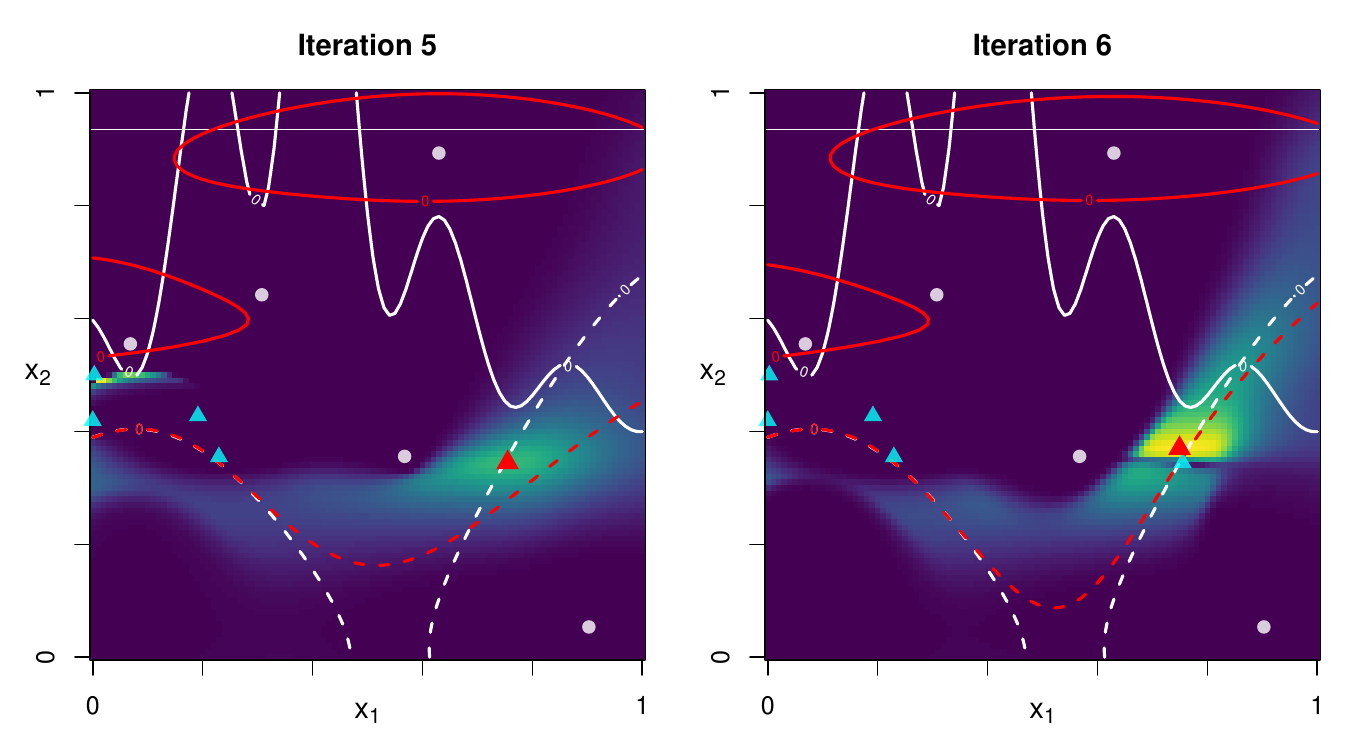}
\caption{Two iterations of exploitation acquisitions for the multimodal
and camelback functions. Heatmaps (yellow/high, purple/low) show $J_n(\mathbf{x}, t_n)$
after 4 jCL acquisitions (left) and 5 jCL acquisitions (right).
White solid/dashed lines show the separate
contours; red solid/dashed lines show the predicted contours. 
White circles indicate an initial LHS, and cyan triangles indicate exploitation 
acquisitions made by our jCL scheme thus far. The red triangle marks the 
next acquisition, where the joint probability is maximized.}
\label{f:joint_step}
\end{figure}

As another visual, Figure \ref{f:joint_step} shows $J_n(\mathbf{x}, t_n)$ in the
full two-dimensional space.  The left panel shows the joint probability after
4 acquisitions ($n=9$ with 5 initial LHS points), resulting in the selection of 
the red triangle.  The right panel shows the updated joint probability after the 
incorporation of the acquired point.  Even though the surrogates are not 
accurately identifying the contours (the red predicted contours are not 
great matches to the white true contours), the acquisition surface which 
incorporates their uncertainty is useful.  Also notice that the strategic shrinking of 
the tolerance ensures that the joint probability will not be optimized 
at the previously observed location. 

\subsection{Exploration} \label{s:exploration}

While the joint probability offers effective exploitation of the joint
contour, sole reliance on it could be futile.  There may be times throughout
the design when the surrogates are not confident that any input would
fall within tolerance for all functions, and the joint probability 
surface may be uninformative.  \added{This could occur due to low accuracy,
high uncertainty, or both. It may also be a sign that no $\tilde{\mathbf{x}}$
exists (more on that momentarily).}  In this case, when we do not have confidence 
in our ability to {\it exploit}, we would rather {\it explore}.

\added{To quantify this intuition,} we set a threshold $p^\star\in(0, 1)$ 
and propose triggering exploratory acquisitions when
$\max_{\mathbf{x} \in \mathcal{X}} \; J_n(\mathbf{x}, t_n) < p^\star$.
\added{The $p^\star$ parameter is another user-controlled tuning parameter indicating
what probabilities would be considered large enough to indicate confidence
in an exploitative acquisition.  Larger $p^\star$ values are more conservative and
will trigger more exploration and less exploitation.
We find $p^\star = 0.2$ works well, and we use this default in all our exercises.
Further comparisons are provided in Supplement \ref{a:tuning}.}
To see this in action, the horizontal grey lines in Figure \ref{f:mm_cb_slice}
mark $p^\star=0.2$.  In the center panels, the joint probability 
does not surpass this threshold, which we take as an indication that 
exploration at this stage would be beneficial.

\begin{figure}[h!]
\centering
\includegraphics[width=1\linewidth]{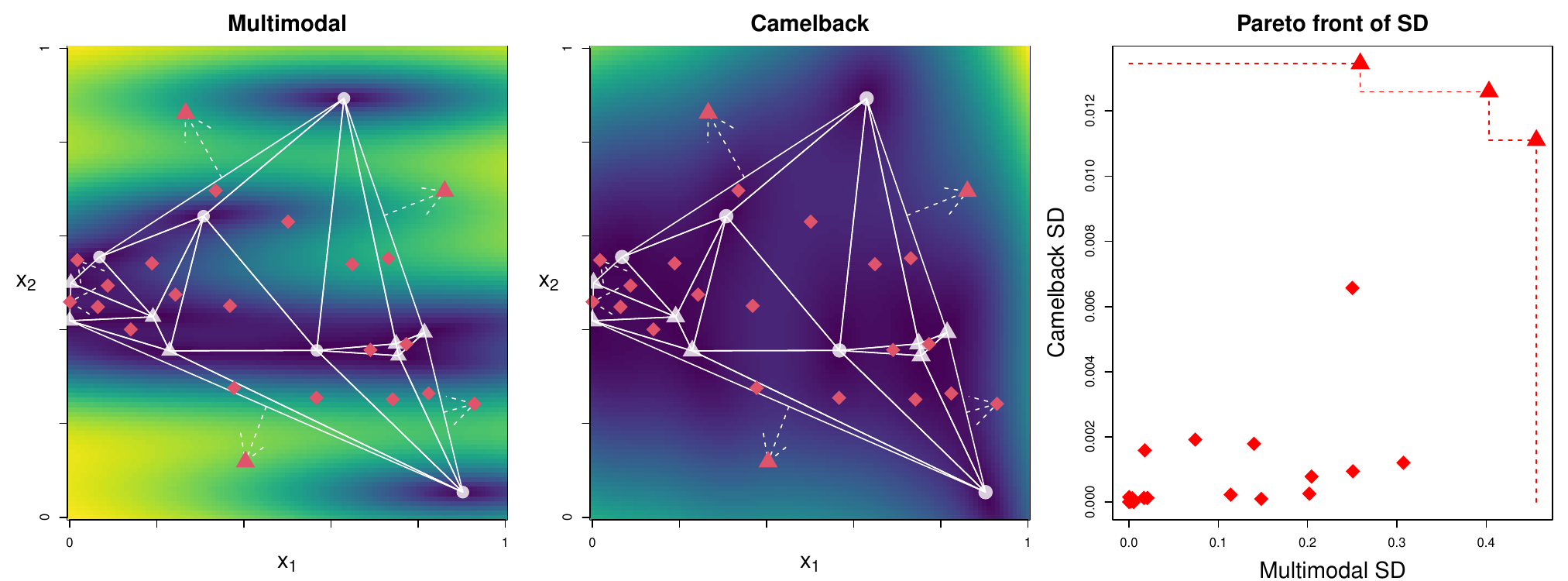}
\caption{{\it Left/Center:} Heatmaps showing posterior predictive standard 
deviations (SD) for the multimodal and camelback functions after 12 observations (5 LHS + 7 jCL).
White lines show Delaunay triangulation of observed locations.  Red points show 
tricands. {\it Right:} SD's for each tricand.  Across all panels, the candidates
on the Pareto front are marked with triangles.}
\label{f:tricands}
\end{figure}

\added{A reasonable exploratory acquisition will increase accuracy and reduce
uncertainty across the surrogates for all functions.  We thus
target locations where surrogate uncertainty is high
for all $r=1,\dots,R$.  This choice is inspired by previous works on sequential
design for variance reduction, such as those leveraging
active learning MacKay, active learning Cohn, and
integrated mean squared error acquisition functions 
\citep[as reviewed in][Chapter 6.2]{gramacy2020surrogates}.} 
Let $\sigma_n^{(r)}(\mathbf{x})$ represent the posterior
standard deviation for the $r^\textrm{th}$ function at input $\mathbf{x}$, given $n$
observed data points.  [For a GP surrogate, this would take the form of $\sigma$
in Eq.~(\ref{e:pred})].  We seek an acquisition that has high $\sigma_n^{(r)}(\mathbf{x})$
for all $r=1,\dots, R$.  It is highly unlikely that all of these standard deviations
would be optimized at the same $\mathbf{x}$, so we instead target their Pareto front.
Let $c(\mathbf{x}) = \{\sigma_n^{(1)}(\mathbf{x}), \cdots, \sigma_n^{(R)}(\mathbf{x})\}$
denote our $R$-dimensional criterion of interest.  An input $\mathbf{x}'$ is said
to ``dominate'' $\mathbf{x}$ if $\sigma_n^{(r)}(\mathbf{x}') \geq \sigma_n^{(r)}(\mathbf{x})
\;\forall\; r=1,\dots,R$.  The Pareto front of $c(\mathbf{x})$ contains
all $\mathbf{x}\in\mathcal{X}$ which are ``non-dominated,'' meaning there are no other
inputs that dominate them, as demonstrated in the right panel of Figure \ref{f:tricands}.

Although the Pareto front is technically a continuum, we prefer to work over a 
discrete set of candidates.  To ensure candidates are placed in regions of high
uncertainty, we use ``triangulation candidates'' \citep[tricands;][]{gramacy2022triangulation}.
To explain, we offer a demonstration in Figure \ref{f:tricands}.  The heatmaps in
the left and center panels show $\sigma_n^{(r)}$ for the multimodal and camelback
surfaces with $n=12$ (continued from previous figures).  We already saw in Figure
\ref{f:mm_cb_slice_b} that the joint probability at this stage did not exceed
$p^\star$, so an exploratory acquisition is warranted.  The white circles in Figure
\ref{f:tricands} represent the observed inputs thus far.  Starting with a 
Delaunay triangulation of the existing locations (solid white lines), tricands 
(red diamonds and triangles) are proposed at the interior of each triangle and 
extending from facets of the convex hull.  By intentionally spreading candidates away
from observed locations, tricands are likely to end up in regions of high uncertainty.

To choose an acquisition from the proposed tricands, we select a tricand
from the Pareto front of $c(\mathbf{x})$ as defined above.  The right panel of 
Figure \ref{f:tricands} visualizes the $c(\mathbf{x})$ criterion across the
proposed tricands, with candidates on the Pareto front as triangles.
Notice how these locations have comparatively high uncertainties for both functions,
as seen in the left/center panels.
The magnitude of each $\sigma_n^{(r)}$ is only relevant within $r$, not across $r$, 
another distinct advantage of the Pareto front strategy.
When there are multiple candidates on the Pareto front, we follow \citet{booth2025contour}
in selecting one at random.

\subsection{Putting it all together}
\label{s:put_together}

Here, we summarize our comprehensive jCL procedure (we will provide specific
implementation details in Section \ref{ss:implement}).  We start by acquiring an 
initial space-filling design of size $n$ and training \added{GP-based} surrogates 
for each function.  Then, a single acquisition 
proceeds as follows.  Set $t_n$ following Eq.~(\ref{e:tol}).  Identify
$\max_{\mathbf{x}\in\mathcal{X}} J_n(\mathbf{x}, t_n)$.  If this maximum
exceeds $p^\star$, then \added{exploit} by acquiring the $\mathbf{x}_{n+1}$ 
that yielded this maximum.
If this maximum does not exceed $p^\star$, then \added{explore} by proposing 
tricands and acquiring the candidate on the Pareto front of 
$\{\sigma_n^{(1)}(\mathbf{x}), \cdots, \sigma_n^{(R)}(\mathbf{x})\}$.
Observe $f^{({r})}(\mathbf{x}_{n+1})$ for each $r$, update the surrogates with this
new data, and repeat until the budget is exhausted or a stopping rule is satisfied.
{\color{blue} In short, exploitation acquisitions are the default, and exploration
is only triggered as a ``fail-safe'' if exploitation does not look promising
(i.e. $\max_{\mathbf{x} \in \mathcal{X}} \; J_n(\mathbf{x}, t_n) < p^\star$).}

Now we finally return to Figure \ref{f:mm_cb}, which shows a complete jCL sequential
design.  Cyan triangles indicate exploitative acquisitions, and cyan squares indicate
explorative acquisitions.  The first acquisitions were near the contours at low
values of $x_1$; these observations helped the surrogates learn that the contours,
although nearing each other, did not intersect in that region.  Then there were 
several exploratory steps, all selecting locations far from others providing useful
information about the general shape of the surfaces.  Ultimately, jCL acquisitions
were able to pinpoint the region of $\tilde{\mathbf{x}}$ (red triangle) and hone
in on it effectively, \added{thanks to the gradual shrinking of the tolerance as
new acquisitions got closer and closer to the optimal design point.}

\subsection{\added{Zero or multiple solutions}}
\label{s:zero}

\begin{figure}[h!]
\centering
\includegraphics[width=0.7\textwidth, trim=0 0 0 30, clip=True]{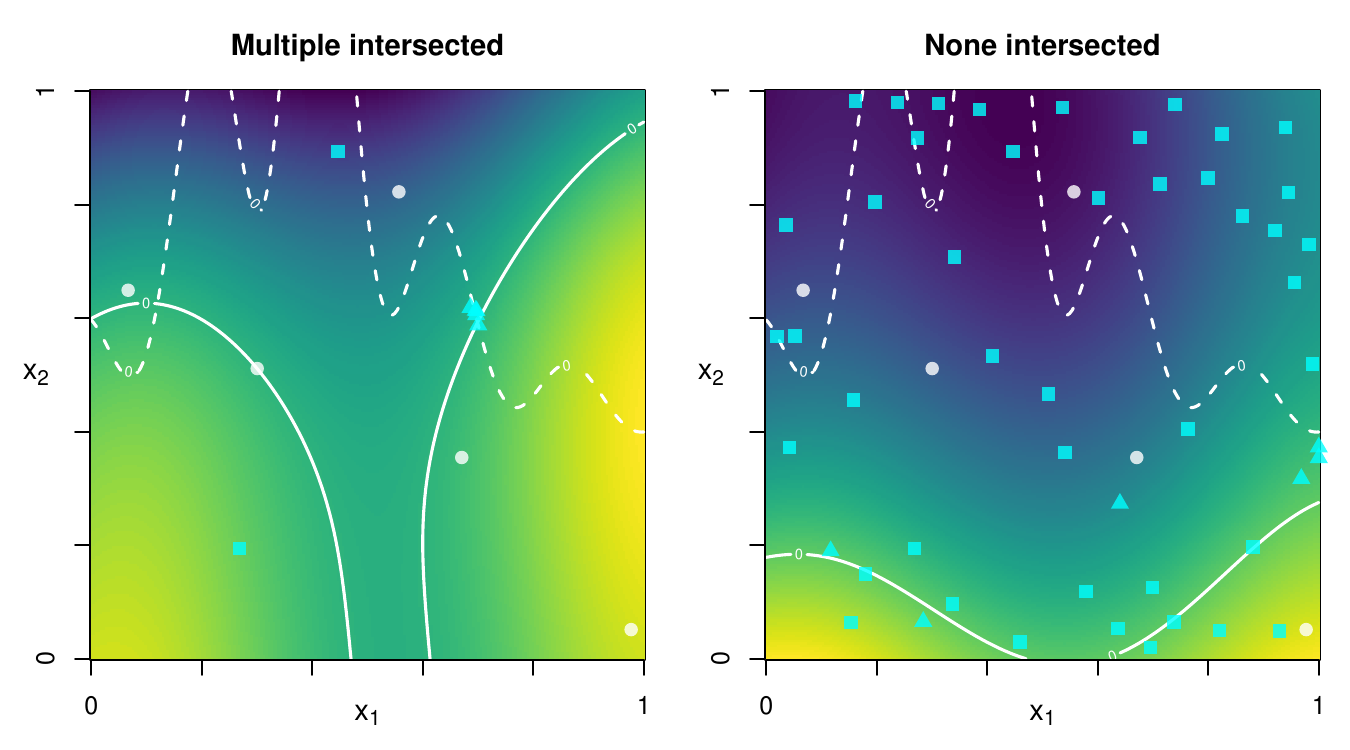}
\caption{\added{Modification of the 2d camelback function so multiple solutions (left)
and no solutions (right) exist.  White lines shows the separate
contours of the multimodal and camelback functions.  White circles indicate an initial LHS,
and cyan squares (exploration) and triangles (exploitation) indicate acquisitions 
made by our jCL scheme. The sampling budget is increased to $50$ for this demonstration.}}
\label{f:multi_none}
\end{figure}

\added{Although our focus is on identifying a single $\tilde{\mathbf{x}}$, it is 
prudent to discuss cases where zero or multiple solutions exist.  Our procedure 
is designed to locate a single solution as efficiently as possible.  If multiple
solutions exist, jCL should find one solution but is not guaranteed to explore 
others.  To demonstrate, we altered the camelback function shown earlier in
Figure \ref{f:mm_cb} so its contour intersects the multimodal function's contour
in three places, as shown in the left panel of Figure \ref{f:multi_none}.  The jCL
design (cyan squares/triangles) finds the rightmost intersection, but then terminates
the design without exploring the other two.  Full identification of all solutions 
will require further methodological development.  We leave this as an interesting 
topic for future work, which we will discuss in Section \ref{s:discuss}.}

\added{Our jCL procedure is well-suited to handle situations where no optimal design
point exists.  In these cases, jCL will tend towards successive explorative 
acquisitions.  If a certain number of exploration steps are triggered in a row 
because no promising directions for exploitation have been found, it is 
reasonable to halt the design.  A sensible value for this number is problem
dependent; it should take into account the severity of missing a solution as 
well as the surface complexity and input dimension (substantial exploration 
may be warranted for complex surfaces).}

\begin{figure}[h!]
\centering
\includegraphics[width=0.7\textwidth]{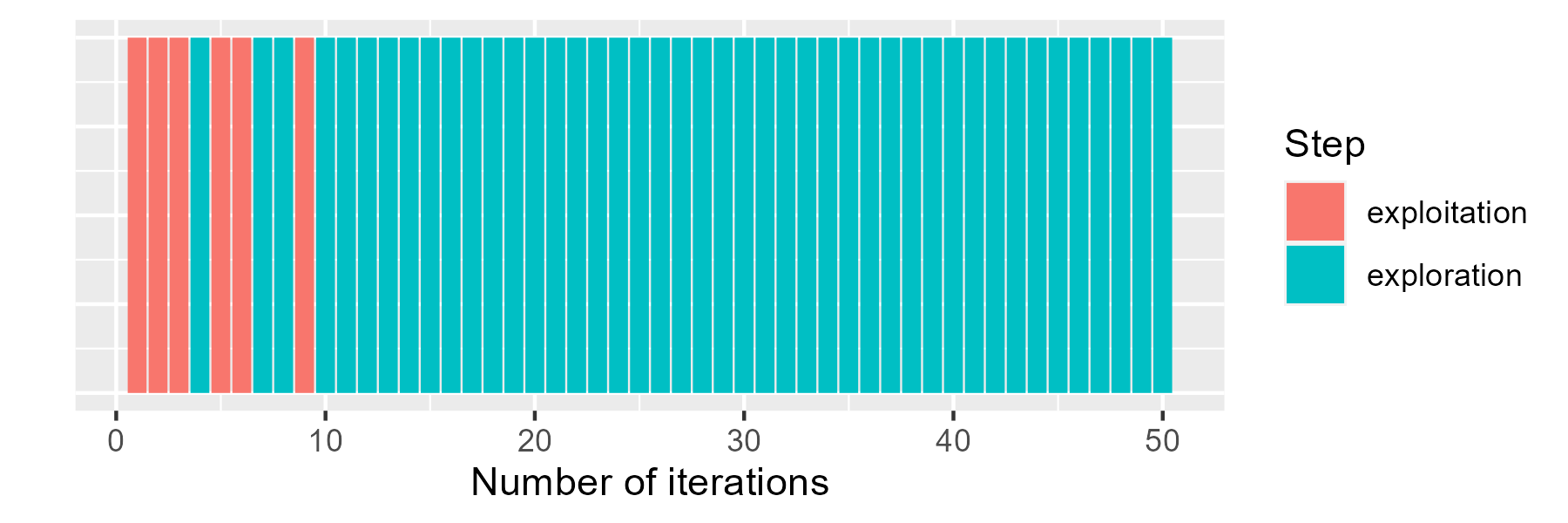}
\caption{\added{Order of acquisitions made for the jCL design shown in the right
panel of Figure \ref{f:multi_none}.}}
\label{f:steps_none}
\end{figure}

\added{To demonstrate, we again altered 
the camelback function, this time so its contour no longer intersects 
with the multimodal function's contour, as shown in the right panel of 
Figure \ref{f:multi_none}.  Cyan squares/triangles again show 
explorative/exploitative acquisitions.  For further context, 
Figure \ref{f:steps_none} shows the order in which these two classes
of acquisitions were made.  The procedure starts with some exploitative
acquisitions where the contours near each other ($x_1\approx 1$, $x_2\approx0.3$),
but the surrogates use these acquisitions to learn that there is actually no
intersection in that region.  By the 10th iteration, every single acquisition
is explorative.  We continue for 50 acquisitions (much longer than necessary) to
show how acquisitions continue to explore without finding any new regions for
exploitation.  The design could reasonably be halted after 10 successive
explorative acquisitions, with confidence that no solution exists.}

\section{Benchmarking}\label{s:sims}

In this section, we will validate jCL's performance against state-of-the-art
alternatives on a variety of benchmark exercises.  We measure performance
using the Euclidean distance between the target $\{\tau_1, \dots, \tau_r\}$ and the
best observed point in $y$-space.  Specifically, denote the ``best squared Euclidean distance''
after $n$ observations as
\begin{equation}\label{e:d}
D_n = \underset{i\in\{1,\dots,n\}}{\min} \; \sum_{r=1}^R (y_{ir} - \tau_r)^2.
\end{equation}
Lower $D_n$ indicates better performance in identifying $\tilde{\mathbf{x}}$ as
defined in Eq.~(\ref{e:objective}).  \added{This performance metric,
akin to simple regret in traditional Bayesian optimization, relies
only on observed data and does not involve the surrogate.  It is possible
to leverage the surrogate to quantify uncertainty in the predicted optimal
design point, but we prefer this ``best observed point'' metric as it 
provides a controlled comparison
across all methods.  With an
eye towards our motivating application, we focus on lower dimensional problems with small
training budgets, but we discuss extensions to larger exercises in Section \ref{s:discuss}.}
Reproducible code for all benchmark exercises is
available in our public git repository.\footnote{
%URL will be provided after blind review.
%Code is provided as a supplemental zip file for review.}
\url{https://bitbucket.org/boothlab/jcl}}

\subsection{Implementation Details}\label{ss:implement}

\added{In this section,} to demonstrate the applicability of our 
procedure, we will deploy both traditional GP and nonstationary DGP 
surrogates.  \added{We train independent surrogates for each function, knowing
that the benchmark functions are indeed independent.} For 
GPs, we use the {\tt GPy} Python-package \citep{gpy2014} with maximum
likelihood estimation of kernel hyperparameters. 
For DGPs, we use the {\tt deepgp} R-package \citep{deepgp} which employs 
ESS sampling of latent layers with Metropolis-Hastings sampling of kernel hyperparameters.
For all surrogates, we use Mat\`{e}rn-5/2 kernels and a fixed nugget of $1.0\times 10^{-6}$ 
to reflect the deterministic nature of each function.  Otherwise, we use software defaults.
We implement all exercises in Python, using the {\tt RPy2} package
\citep{rpy2} to access the {\tt deepgp} functions within Python.

To optimize the joint probability of Eq.~(\ref{e:jointprob}), we employ a multi-start 
numerical optimization using the \texttt{minimize} function from the 
\texttt{SciPy} Python-package \citep{virtanen2020scipy}.  
Multi-start initializations
are seeded with the best $10d$ out of $10,000d$ random points, the optimal 
joint probability point from the previous iteration, and an additional $10d$ 
random points.  We also use a logarithmic trick to avoid instability in the joint probability 
calculation when the tolerance is small.  Instead of optimizing 
$J_n(\mathbf{x}, t_n)$ directly, we optimize
\[
\begin{aligned}
\sum_{r=1}^R \left\{ \ln \bigg( \mathbb{P}\left( \tau_r - t \le f^{(r)}_n(\mathbf{x}) \le \tau_r + t\right) \bigg) \right\}
    &= \sum_{r=1}^R \left\{ \ln \bigg( \underbrace{\mathbb{P}(f^{(r)}_n(\mathbf{x}) \le \tau_r + t)}_{=:a} - 
    \underbrace{\mathbb{P}(f^{(r)}_n(\mathbf{x}) \le \tau_r - t)}_{=:b} \bigg) \right\} \\
    &= \sum_{r=1}^R \left\{ \ln(a) + \ln \bigg(1- \exp \big( \ln(b) - \ln(a) \big) \bigg) \right\}.
\end{aligned} 
\]
We use the {\tt SciPy} Python-package \citep{virtanen2020scipy} to calculate Delaunay 
triangulations and propose tricands following \citet{gramacy2022triangulation}.

In the following exercises, we halt the sequential design when either the total budget
of evaluations has been spent or an acceptable $D_n<\epsilon$ has been reached.
Throughout, we use $\epsilon = 0.001$ after appropriately scaling the responses (scaling 
details are provided in Supplement \ref{a:function}).

\subsection{Competitors} \label{s:competitors}

We consider \added{six} alternative designs (within each exercise, surrogate choices are
kept consistent, the only variations are the methods for selecting training data).
First, we benchmark our jCL design against a space-filling Latin hypercube sample 
of equivalent size.  Although we will report LHS performance as the sample
size is incremented, the LHS design is not sequential and does not use
the surrogates to select design locations.  Next, we compare to two state-of-the-art
contour location schemes: that of \citet{cole2023entropy} and \citet{booth2025contour}.
\citeauthor{cole2023entropy} acquire points based on a local optimization of 
the surrogate's classification entropy, which can be heavy on exploitation.  
\citeauthor{booth2025contour} encourage more exploration by acquiring triangulation
candidates on the Pareto front of entropy and uncertainty.  Both of these
methods have been shown to excel at contour location, but are designed for a 
single function.  To suit them to our setting,
we simply alternate between CL acquisitions for each function.  We refer to
these methods as ``alternating entropy'' and ``alternating Pareto,'' respectively.

\added{Next, we compare against two common optimization algorithms: Bayesian optimization
(``BO'') and zero-finding (``Zero'').  To leverage these methods, we transform the 
problem of Eq.~(\ref{e:objective}) into the following optimization problem:
\[
\tilde{\mathbf{x}} = \underset{\mathbf{x}\in\mathcal{X}}{\mathrm{argmin}} \; h(\mathbf{x})
\quad\textrm{where}\quad
h(\mathbf{x}) = \sum_{r=1}^R \left(f^{(r)}(\mathbf{x}) - \tau_r\right)^2.
\]
The BO procedure trains a single GP/DGP 
surrogate on observed $\{\mathbf{x}_i, h(\mathbf{x}_i)\}$ for $i=1,\dots,n$ then uses the expected
improvement criterion \citep{jones1998efficient} to select acquisitions in order
to minimize $h(\mathbf{x})$.  We use the ``L-BFGS-B'' algorithm \citep{byrd1995limited}
with $10d$ randomly seeded multi-starts to optimize the EI surface for each acquisition.
The zero-finding procedure uses L-BFGS-B directly on the black-box $h(\mathbf{x})$ function
without any surrogate.  When assessing performance, we carefully track the number of
black-box evaluations used by the zero-finding method in order to match it appropriately
to the sequential surrogate-informed methods.}

\added{Lastly, we implement a joint acquisition function inspired by 
the ``targeted adaptive design'' procedure of \citet{graziani2024targeted}, but without
the extra machinery they built-in to accommodate noisy simulations and wider target ranges. 
This method, which we label ``PDF,'' selects acquisitions by optimizing the surrogates' joint 
probability density function centered at the target $(\tau_1, \dots, \tau_r)$.  
For independent GP surrogates with posterior means 
$\mu_n^{(r)}(\mathbf{x})$ and 
posterior standard deviations $\sigma_n^{(r)}(\mathbf{x})$, acquisitions are chosen 
following
\[
\mathbf{x}_{n+1} = \underset{\mathbf{x}\in\mathcal{X}}{\mathrm{argmin}} \; g(\mathbf{x})
\quad\textrm{where}\quad
g(\mathbf{x}) = \prod_{r=1}^R 
\phi\left(\frac{\tau_r - \mu_n^{(r)}(\mathbf{x})}{\sigma_n^{(r)}(\mathbf{x})} \right),
\]
and $\phi$ represents the standard Gaussian PDF.  This acquisition function is
similar to our joint probability metric, but it omits the shrinking tolerance $t_n$
and loses the interpretability of the probability, which is key to informing our
exploratory acquisitions.
To maintain fair comparisons, we optimize the PDF acquisition
function with the same optimization procedures used for the aforementioned methods.}

\subsection{\added{Results}}
\label{s:synthetic}

We consider three test cases.  In each setting, we prescale inputs to the unit
cube, i.e., $\mathcal{X} = [0, 1]^d$, and prescale responses to have
unit variance.  We use $\tau_r = 0$ for all $r$, after shifting the 
functions so there exists a single $\tilde{\mathbf{x}}$ (details
provided in Supplement \ref{a:function}).
First, we continue the 2d multimodal and camelback exercise
of Figures \ref{f:mm_cb}--\ref{f:tricands} (with $d=R=2$).  We use 
GP surrogates with an initial LHS of size $n_0 = 5$ and a total budget 
of $n = 25$.  Second, we use two variations of the 2d ``Gramacy'' function 
\citep{gramacy2009adaptive}, which is characterized by large flat regions
with a single ``hill'' and ``valley.''  A visual of these surfaces
and one of our jCL designs is provided in Supplement \ref{a:gramacy}.  To 
accommodate the nonstationarity
of the surfaces, we use DGP surrogates, again with $d=R=2$, $n_0=5$, and $n=25$.
Finally, we expand to $d=R=3$ with an adaptation of the multimodal function,
the Ishigami function \citep{ishigami1990importance}, and a third ``trig'' function 
consisting of several sine and cosine functions.  These surfaces are relatively
stationary, so we return to GP surrogates. For this larger 
dimension, we start with an LHS of size
$n_0=10$ and acquire up to $n=40$.  We repeat each exercise for 50 Monte Carlo
repetitions with re-randomized starting designs.  

\begin{figure}[ht!]
\centering
\includegraphics[width=1\linewidth, trim=135 650 80 550, clip]{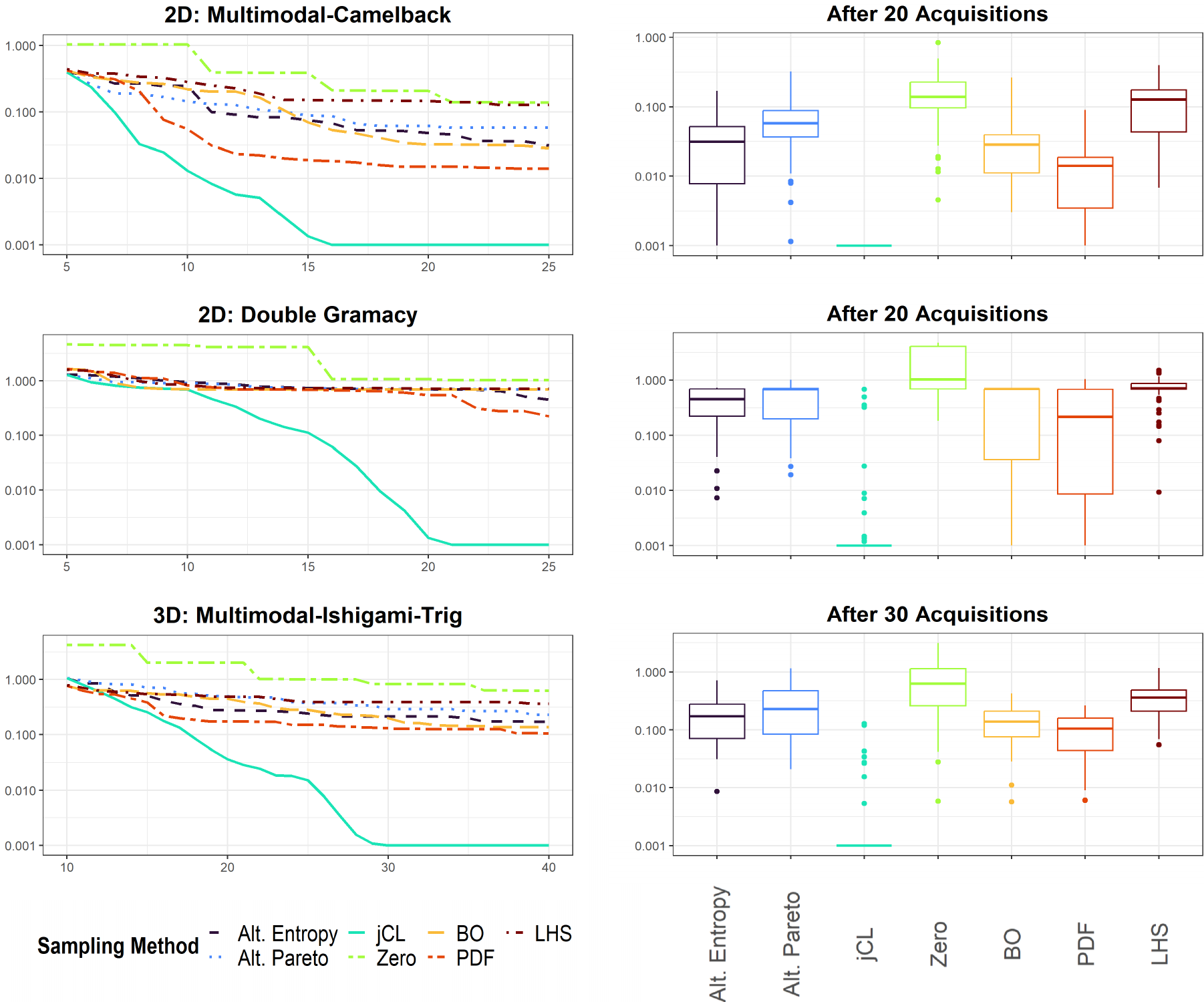}
\caption{\added{Performance in $D_n$ (Eq.~\ref{e:d}) on the log scale. Line plots 
show median performance as $n$ is incremented. Box plots show the 
spread at the end of sequential sampling.  Top and bottom panels use GP surrogates;
center panel uses DGP surrogates.}}
\label{f:all_results}
\end{figure}

Performance in $D_n$ is shown in Figure \ref{f:all_results}.  Our jCL design 
\added{(solid teal)} consistently outperforms the competitors.  In all three 
examples, jCL has a much steeper decline in $D_n$.
For the 2d multimodal and camelback functions, 
jCL achieves better median performance after 5 acquisitions than the other 
methods could achieve after $20$. For the other examples, after about $10$
acquisitions jCL achieves median performance that is lower than the
``almost best'' performance of the other methods. 
Our jCL design is the only one to consistently reach the specified
$\epsilon$, and it often does so before spending the maximum allowed budget.
\added{The alternating contour location methods struggle as they waste 
observations learning other non-intersecting regions of the contours.  Of
the optimization methods, BO which makes careful use of a surrogate model
outperforms the zero-finding optimization algorithm, but it still pales in
comparison to jCL.  We suspect this is largely due to the complexity of the
$h(\mathbf{x})$ surface compared to the separate $f^{(r)}(\mathbf{x})$ surfaces.  The 
$h(\mathbf{x})$ is undoubtedly less smooth, more complex, and more 
challenging to learn than the separate functions.  The PDF method is
second-best, but it still struggles compared to jCL.  Although the PDF
acquisition function involves surrogate uncertainty, it does not have a strong
mechanism for exploration.  Its efforts to exploit can also be less
aggressive than those of the joint probability thanks to the shrinking tolerance.}

\section{High-Speed Army Reference Vehicle}\label{s:harv}

\begin{figure}[h!]
\centering
\includegraphics[width=.4\linewidth]{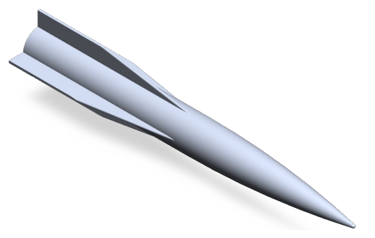}
\includegraphics[width=.5\linewidth, trim=0 0 0 20, clip]{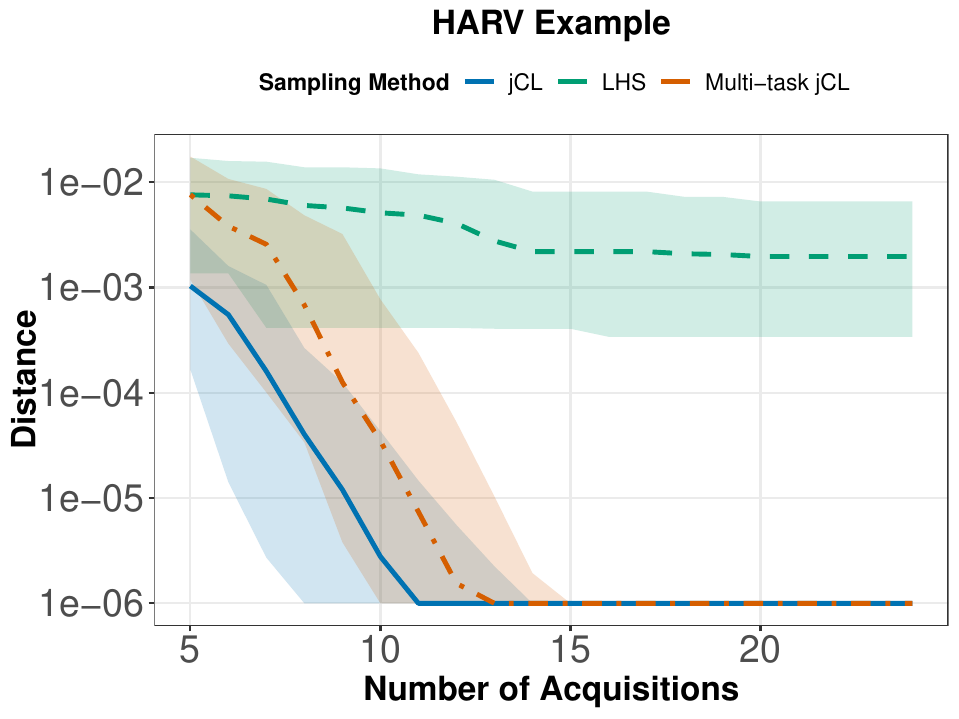}
\caption{{\it Left:} HARV geometry featuring four deflectable fins. {\it Right:} 
Performance in $D_n$ (Eq.~\ref{e:d}) across acquisitions.}
\label{f:data}
\end{figure}

Here we consider our motivating application: CFD simulations
of the rotational torques acting on the high-speed army reference
vehicle across various free-stream conditions.  We use NASA's CBAERO 
(which is preferred for its 
computation speed) in this study to generate numerical 
approximations of the forces and moments on the vehicle \citep{kinney2007aerothermal}. 
The left panel of Figure \ref{f:data} shows
the HARV geometry.  The deflection of the fins are the key controllable
inputs.  We seek the optimal deflection of the fins
($\tilde{\mathbf{x}}$) that will result in stable flight for given
flight conditions (Mach, angle of attack, etc.), with as few CBAERO
evaluations as possible.  Here, we focus on the 
configuration of the two side fins, targeting zero pitch moment
and zero roll moment (we fix the top and bottom fins and the sideslip 
angle to focus on the longitudinal degrees of freedom).

\added{Since there is reason to suspect correlation between pitch and roll,
we deploy our jCL procedure with both independent GP surrogates and a
multitask GP surrogate \citep[implemented in {\tt botorch};][]{balandat2020botorch}.
We use an initial design of size $n_0 = 5$, a maximum budget of 
$n = 25$, and a stopping criterion of $\epsilon = 1\times 10^{-6}$.
We compared performance in $D_n$, omitting the poorer performers from
Figure \ref{f:all_results} but retaining the LHS as a baseline.}
Performance across 50 repetitions is shown
in the right panel of Figure \ref{f:data}. Our jCL designs were able
to consistently locate the optimal fin configuration with fewer than 15
observations. \added{Curiously, the multitask GP slightly underperforms
the independent GPs, but still quickly reaches the target accuracy.
The space-filling designs made little-to-no progress.
Finding the optimal wing configuration with only 10-15 observations enables
us to repeat this procedure across a variety of flight conditions and top/bottom
wing configurations in order to build out a trimmed aerodynamic database.}

\section{Conclusion}\label{s:discuss}

We have proposed a novel joint contour location (jCL) sequential design to
identify the optimal design point from multiple computer experiments
(Eq.~\ref{e:objective}).  Our method uses the joint probability of the multiple
responses being within a specified, converging tolerance in order to exploit.
It uses triangulation candidates on the Pareto front of each surrogate's posterior
standard deviation as a ``failsafe'' in order to explore when exploitation is not advised.
Through careful, strategic design, jCL is able to outperform standard CL 
\added{and optimization-based designs} with far fewer observations of the 
black-box computer experiments.
Our jCL is also versatile; it can be implemented using any surrogate that
provides posterior predictive distributions.

\added{In this work we targeted lower dimensional settings ($d\leq 3$)
with small evaluation budgets ($n\leq 40$).  While our jCL procedure
is applicable to larger scenarios, some modifications could be beneficial
in higher dimensions.  The triangulation candidates used in 
our exploration step may become cumbersome and ineffective for large $d$.
Adaptations such as Voronoi candidates \citep{wycoff2024voronoi} could
be advantageous.  Furthermore, the Pareto front could become more diffuse and
less informative as dimension grows.  Modified strategies for exploratory 
acquisitions that still balance uncertainty across all functions may be
warranted.  Furthermore, we assumed that each $f^{(r)}$ would be evaluated
together for the same input $\mathbf{x}$, which we treated as a single evaluation
of the black-box function.  If each $f^{(r)}$ is evaluated separately,
then additional methodology to select which black-box
function to query at each acquisition would be beneficial.}

\added{If multiple solutions exist (either at a discrete number of locations
or on a continuum), our jCL procedure is designed to locate a single optimal
design point efficiently.  It will not ensure exploration of other solutions.
We suspect strategic thresholds on the joint probability and adjustments to the
tolerance refinement would facilitate identification of multiple solutions,
but this would require careful investigation.
For the HARV geometry, methods to identify multiple optima could facilitate
application of jCL over the full parameter space, rather than for fixed
flight conditions.}

\section*{Acknowledgments} This work was performed under the auspices of 
the U.S. Department of Energy by
Lawrence Livermore National Laboratory under Contract DE-AC52-07NA27344,
LLNL-JRNL-2013503-DRAFT. The authors thank Brad Perfect for his invaluable suggestions and feedback. 

\section*{Data Availability} Reproducible code including data generation for all 
synthetic exercises is provided
in a public git repository
%For blind review, codes are provided as a supplemental 
%zip file.
at \url{https://bitbucket.org/boothlab/jcl}.

\singlespacing
\bibliography{main}
\bibliographystyle{jasa}

\newpage
\appendix
\begin{center}{\Large\bf SUPPLEMENTARY MATERIAL}
\end{center}

\section{Synthetic Functions} \label{a:function}

In this section we provide details for the three test functions used in
Section \ref{s:sims}.  We have adjusted and scaled each function to ensure
that a single $\tilde{\mathbf{x}}$ exists at $\tau_r = 0$ for all 
$r\in\{1,\dots,R\}$.

\subsubsection*{2D: Multimodal-Camelback} 
The ``multimodal'' function \citep{bichon2008efficient} is defined as
\vspace{2mm}
\[
y_1 = \frac{\frac{x_2 - 1}{20} \left(x_1^2 + 4\right) - \sin\left(\frac{5x_1}{2}\right) - 2}{3.556} - 0.00678656
\quad\textrm{for}\quad x_1\in[-4, 7],\;\; x_2\in[-3,8].
\]
\vspace{2mm}

\noindent The ``camelback'' function \citep{molga2005test} is defined as
\vspace{2mm}
\[
y_2 = \frac{u^2 \left(4 - 2.1 u^2 + \frac{u^4}{3}\right) + 2 uv + \left(\frac{26}{9} v^2 \left(-4 + \frac{16v^2}{9}\right)\right) - 0.1 }{2.242} - 0.00719266
\quad\textrm{for}\quad x_1\in[-1, 1],\;\; x_2\in[0,1],
\]
\vspace{2mm}

\noindent where $u =1.2 x_1 - 0.1$ and $v = 0.9 x_2$.

\subsubsection*{2D: Double Gramacy} 
The ``double Gramacy'' functions \citep{gramacy2009adaptive} are defined as 
\vspace{2mm}
\[
\begin{aligned} 
y_1 &= 9.27 \left[x_1 \cdot \exp\left(\left(-x_1-0.5\right)^2 - \left(x_2+0.5\right)^2\right) - 0.1\right] + 0.09830494 \\ 
y_2 &= \frac{\left(x_2 + 0.5\right) \exp\left(-\left(\frac{x_2}{4}\right)^2 - x_1^2\right) - 1}{0.4975} - 0.01300846 
\end{aligned}
\]
\vspace{2mm}

\noindent For both functions, the domain is $x_i\in[-2,6]$ for $i\in\{1, 2\}$.

\subsubsection*{3D: Multimodal-Ishigami-Trig} 
The 3-dimensional ``multimodal'' function is defined as
\vspace{2mm}
\[
y_1 = \frac{\frac{x_2 - 1}{20} \left(x_1^2 + 4\right) - \sin\left(\frac{5x_1}{2}\right) - 2 + x_3 - 0.00052352}{3.588}
\quad\textrm{for}\quad x_1\in[-4, 7],\;\; x_2\in[-3,8],\;\;x_3\in[0,1].
\]
\vspace{2mm}

\noindent The ``Ishigami'' function \citep{ishigami1990importance} is defined as
\vspace{2mm}
\[
y_2 = \frac{ \sin\left(x_1\right) + 7 \sin\left(x_2\right)^2 + 0.1  x_3^4 \sin\left(x_1\right) + 2.79921514 }{3.72}
\quad\textrm{for}\quad x_i\in[-\pi, \pi] \;\forall\; i\in\{1,\dots,3\}.
\]
\vspace{2mm}

\noindent The ``Trig'' function is defined as
\vspace{2mm}
\[
y_3 = \frac{ \sin\left(x_1\right) + \cos\left(x_1\right) + x_2^2 + \sqrt{x_3} + \sin\left(x_3\right) - 3.05174287 }{0.582}
\quad\textrm{for}\quad x_i\in[0, 1] \;\forall\; i\in\{1,\dots,3\}.
\]

\section{Double Gramacy Design}\label{a:gramacy}

\begin{figure}[ht!]
\centering
\includegraphics[width=0.8\textwidth]{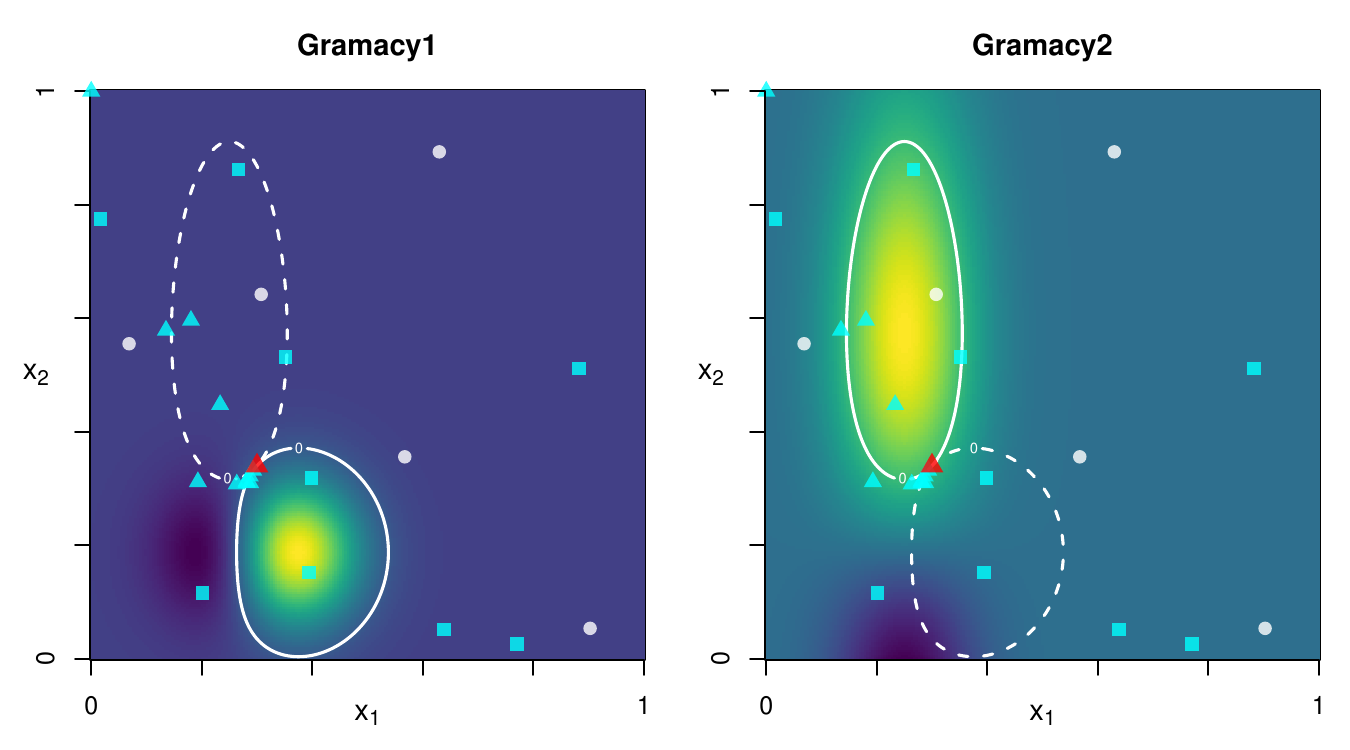}
\caption{
Heatmaps (yellow/high, purple/low) of the 2d Gramacy functions
\citep[derived from][]{gramacy2009adaptive}.  White lines shows the separate
contours $f^{(r)}(\mathbf{x})=0$, and red triangle marks the optimal design
point where both contours are zero.  White circles indicate an initial LHS,
and cyan squares (exploration) and triangles (exploitation) indicate acquisitions 
made by our jCL scheme.}
\label{f:dg}
\end{figure}

Figure \ref{f:dg} shows the two surfaces of the ``double Gramacy'' 
example, with contours at $\tau_1 = \tau_2 = 0$ in solid/dashed white.  
Cyan points show a jCL design with a DGP surrogate.  The combination of 
exploitation steps (triangles) and exploration steps (squares) effectively
locates the optimal design point (red triangle).

\section{Examining tuning parameters}
\label{a:tuning}

\begin{figure}[h!]
\centering
\subfigure[\added{Varying values of $p^*$ (using $w=0.9$)}]{\includegraphics[width=0.49\textwidth]{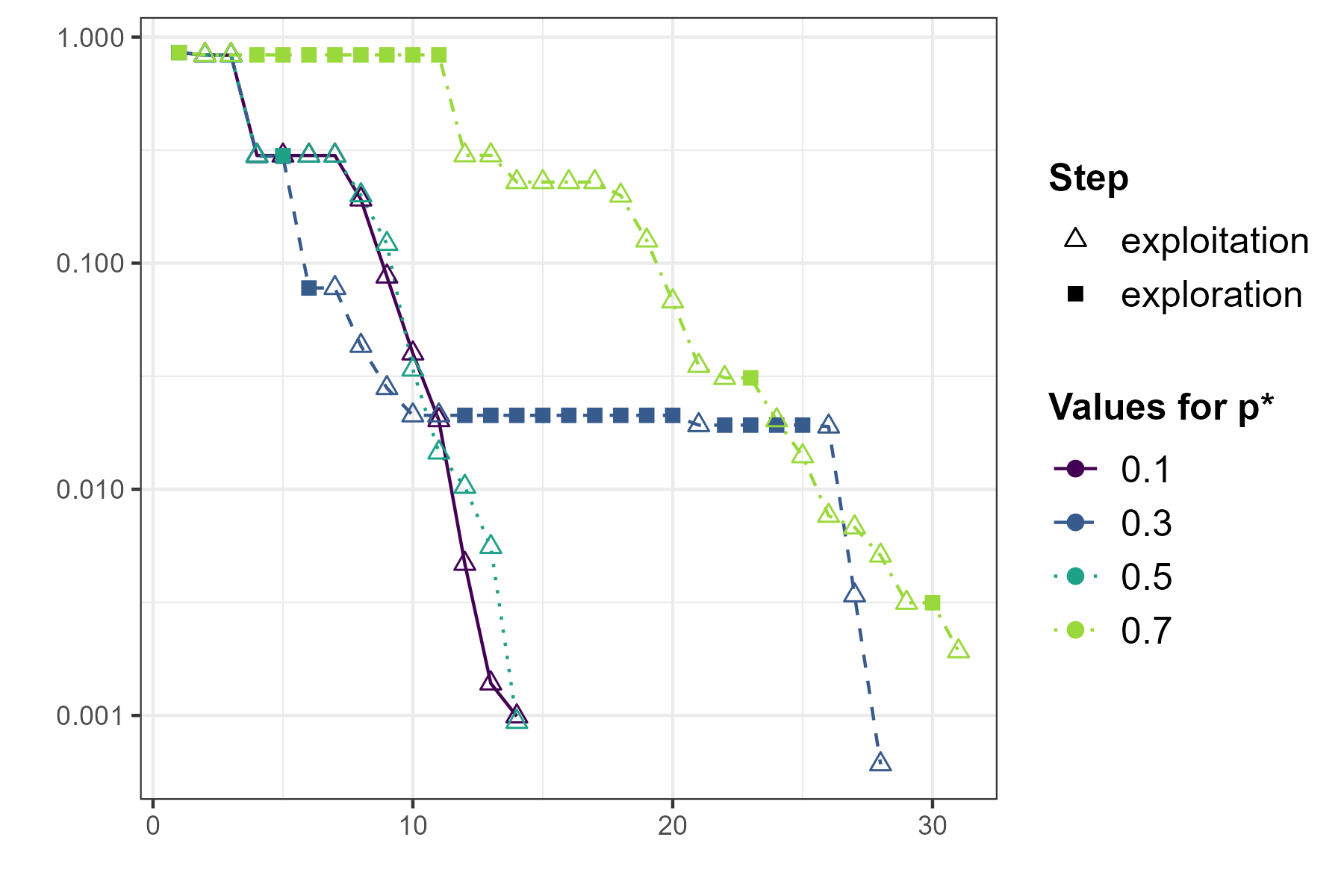}}
\subfigure[\added{Varying values of $w$ (using $p^\star=0.2$)}]{\includegraphics[width=0.49\textwidth]{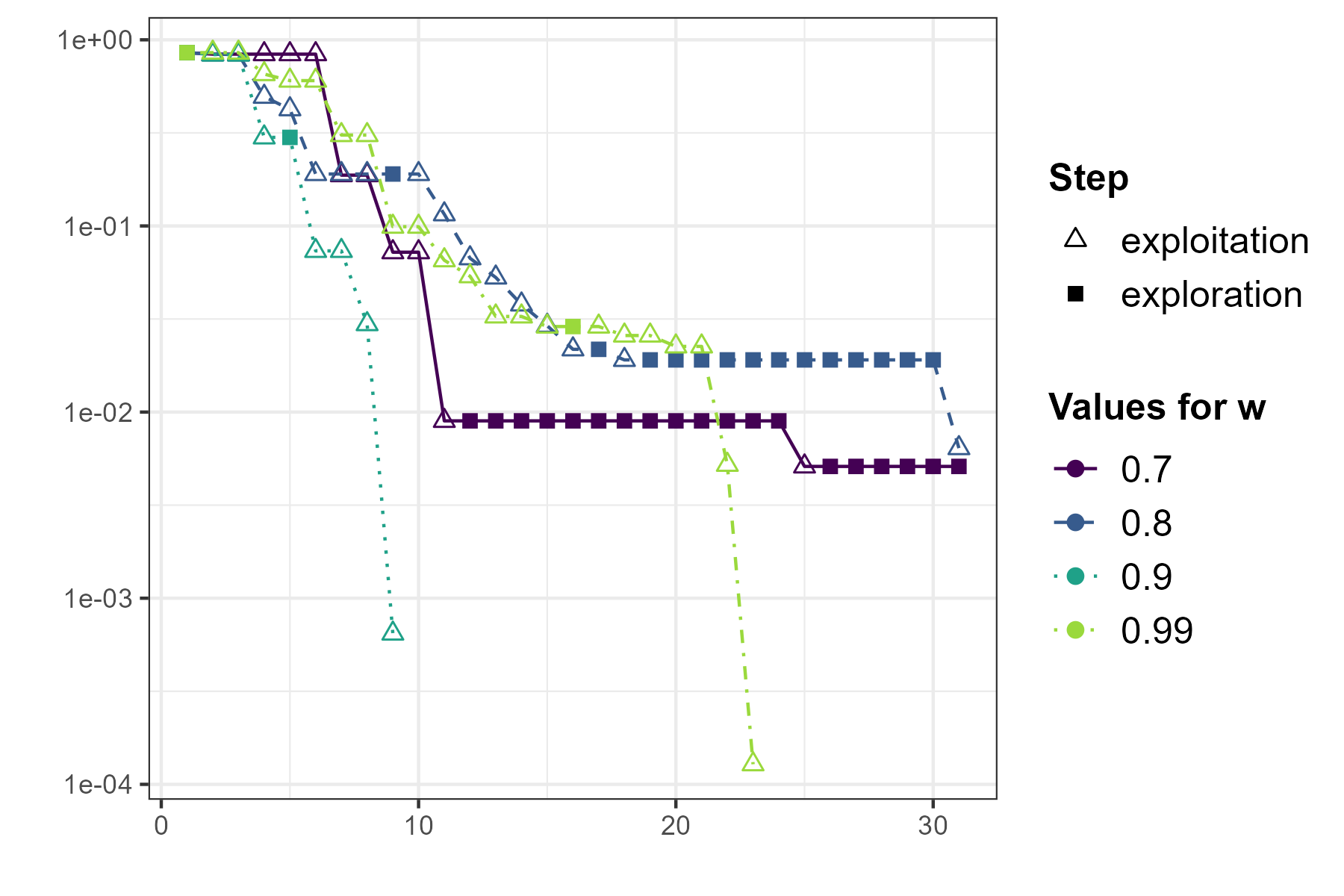}}
\caption{\added{Performance in $D_n$ for jCL designs of the 3D Multimodal-Ishigami-Trig example
with varying tuning parameter values.  All designs start with the same initial LHS.  Squares indicate 
exploration steps, and triangles indicate exploitation steps.}}
\label{f:p_star_w}
\end{figure}

\added{To examine the effect of $p^\star$ and $w$ on our jCL procedure, we revisit
the 3D Multimodal-Ishigami-Trig example of Section \ref{s:sims}, this time varying
the values of $p^\star$ and $w$ while keeping the starting design constant.
The left panel of Figure \ref{f:p_star_w} shows performance in $D_n$ for 
$p^\star\in\{0.1, 0.3, 0.5, 0.7\}$.  Larger $p^\star$ values
encourage more exploration.  This is apparent from the design with $p^\star=0.7$, which
starts with many exploratory steps without making any progress.  Eventually, the procedure
will find an avenue for exploitation, but it may have wasted acquisitions in the process.
There is variability in performance across lower $p^\star$ values caused by the unique
acquisitions made by each design (with $p^\star = 0.1$ and $p^\star = 0.5$ performing similarly).
Ultimately, we landed on a default of $p^\star=0.2$, which will 
avoid excessive exploration without discouraging it altogether, but we believe the interpretable
and user-adjustable nature of this tuning parameter is a strength.}

\added{Next, we examine how the value of $w$ affects jCL's performance. The tuning 
parameter $w$ controls how much the tolerance $t_n$ shrinks as each ``better''
design point is observed. A smaller value of $w$ shrinks the tolerance more aggressively, 
and a larger value shrinks the tolerance more gradually.  The only restriction is that $w$
must be strictly less than 1.  The right panel of Figure \ref{f:p_star_w} shows 
results on the same 3D example with a variety of $w$ values.  Small values which are
attempting to be more aggressive actually have the opposite effect; they shrink the tolerance
too much, reducing the maximum joint probability and resulting in over-exploration.  
We caution against setting $w$ too low and recommend $w=0.9$ as a default.}

\end{document}